\documentclass[twocolumn,superscriptaddress,amsmath,aps]{revtex4}
\usepackage{graphicx,color}
\usepackage{bm}
\usepackage[hypertex]{hyperref}

\newcommand{\be}{\begin{equation}}
\newcommand{\ee}{\end{equation}}
\newcommand{\bea}{\begin{eqnarray}}
\newcommand{\eea}{\end{eqnarray}}
\newcommand{\bsube}{\begin{subequations}}
\newcommand{\esube}{\end{subequations}}

\newcommand{\Eq}[1]{Eq.\,(\ref{#1})}
\newcommand{\Eqs}[1]{Eqs.\,(\ref{#1})}

\newcommand{\dg}{\dagger}
\newcommand{\la}{\langle}
\newcommand{\ra}{\rangle}

\newcommand{\nl}{\nonumber \\}

\newcommand{\gam}{\gamma}

\newcommand{\beq}{\begin{equation}}
\newcommand{\eeq}{\end{equation}}
\newcommand{\beqn}{\begin{eqnarray}}
\newcommand{\eeqn}{\end{eqnarray}}
\newcommand{\bsub}{\begin{subequations}}
\newcommand{\esub}{\end{subequations}}

\newcommand{\fdg}{f^\dagger}

\begin{document}

\title{Majorana Conductances in Three-Terminal Transports}

\author{Xin-Qi Li}
\email{xinqi.li@tju.edu.cn}
\affiliation{Center for Joint Quantum Studies and Department of Physics,
School of Science, Tianjin University, Tianjin 300072, China}
\author{Wei Feng}
\email{fwphy@tju.edu.cn}
\affiliation{Center for Joint Quantum Studies and Department of Physics,
School of Science, Tianjin University, Tianjin 300072, China}
\author{Lupei Qin}
\affiliation{Center for Joint Quantum Studies and Department of Physics,
School of Science, Tianjin University, Tianjin 300072, China}
\author{Jinshuang Jin}
\affiliation{Department of Physics, Hangzhou Normal University,
Hangzhou, Zhejiang 310036, China}

\date{\today}

\begin{abstract}
We consider a two-lead (three-terminal) setup of nonlocal transport
through Majorana zero modes (MZMs) and construct
a Majorana master equation (which is also valid for small bias voltage).
We first carry out representative results of current and
then show that a modified Bogoliubov-de Gennes (BdG) treatment
can consistently recover the same results.
Based on the interplay of the two approaches,
we reveal the existence of nonvanishing channels
of teleportation and crossed Andreev reflections
even at the limit $\epsilon_M\to 0$ (zero coupling energy of the MZMs),
which leads to new predictions for
the height of the zero-bias-peak of the local conductance
and the $\epsilon_M$-scaling behavior of the teleportation conductance,
for verification by experiments.
\end{abstract}


\maketitle


Majorana fermions obey non-Abelian statistics and have sound potential
for topological quantum computations \cite{Kita01,Kit03,Sar15}.
For the purpose of identification, the self-Hermitian property and nonlocal nature
of the Majorana zero modes (MZMs) indicate some unique transport phenomena such as
fractional Josephson effects \cite{Sar10,Opp10,Fu09,Cay17,JE-1,JE-2},
peculiar noise behaviors \cite{Dem07,Bee08,Zoch13,Shen12,Li12,Law09},
and resonant Andreev reflections (AR) \cite{Law09,BNK11}
which also result in the zero-bias peak
of conductance and a quantized height of $2e^2/h$
\cite{Law09,BNK11,Sarma01,Flen10,Flen16,Kou12,Kou19}.
Recent interest also includes the nonlocal transport signatures
\cite{Gla16,DS17d,Mar18b,Mar18c,Sch-1,Sch-2,BCS-1a,BCS-1,BCS-2}
which may help to distinguish the nonlocal MZMs
from the topologically trivial Andreev states
\cite{DS17,Agu17,Cay19,Agu18,abs-1,abs-2,abs-3,abs-4,Vu18}.

The genuinely nonlocal nature of the MZMs should be associated with
such as the teleportation \cite{Sem06ab,Lee08,Fu10}
or the crossed correlation of two remote majoranas
($\gamma_1$ and $\gamma_2$) \cite{Dem07,Bee08,Zoch13,Li12,Law09}.
However, based on the usual single-lead local measurements,
either the zero-bias peak (ZBP) or the quantized conductance $2e^2/h$
has been regarded not sufficient to conclude the demonstration of the MZMs.
Therefore, nonlocal transport through a two-lead setup,
which is actually a three-terminal device
(with two normal leads coupled to a grounded superconducting terminal),
can be considered as a more powerful platform \cite{Sch-1,Sch-2,BCS-1a,BCS-1,BCS-2},
in particular for the purpose to demonstrate the Majorana nonlocality
via such as teleportation and/or crossed AR (CAR) evidence.
In Ref.\ \cite{Bee08}, it was found that the CAR channel dominantly suppresses
the local AR (LAR) contribution under the limit $\epsilon_M\gg (eV, \Gamma_i)$,
i.e., the Majorana coupling energy $\epsilon_M$ much larger than
the equally biased voltage $V$ and the coupling rate to the leads $\Gamma_i$.

However, at the opposite limit $\epsilon_M\to 0$, it was found that
the cross correlation of currents in the two leads vanishes \cite{Dem07,Bee08,Zoch13}.
In the Bogoliubov-de Gennes (BdG) scattering approach treatment \cite{Bee08,Zoch13},
the zero cross correlation of currents is rooted in
the vanishing teleportation and CAR channles at the limit $\epsilon_M\to 0$,
in terms of a picture of disconnected MZMs or, equivalently,
destructive interference between the `positive' and `negative' energy states.
Nevertheless, if applying a {\it BdG free} treatment in terms of the MZMs associated
regular fermion occupation-number-states \cite{Li20,Li21a,Li21b},
both channels of the teleportation transfer and the CAR process
naturally exist there, without much difference for $\epsilon_M\to 0$ or not.
Despite that the BdG-free occupation-number-state treatment can also
result in the zero cross correlation of currents,
the underling reason is different and is owing to
a ``degeneracy" of the teleportation and the AR process channels \cite{Li21a,Li21b}.  

In this work, we revisit the two-lead (three-terminal) transport setup
considered in Ref.\ \cite{Bee08}
and construct a Majorana master equation (MME) which,
beyond the limitation under the Born-Markov approximation,
is also applicable for small bias voltage.
Based on the MME, we first carry out the representative results of current and then
show that a modified BdG treatment can consistently recover the same results.
We further carry out new predictions
for the Majorana conductances associated with the three-terminal device. \\
\\
{\it Majorana Master Equation}.---
The low-energy effective Hamiltonian for a topological superconductor (TS) wire
hosting a pair of MZMs
can be commonly formulated as $H_M=i\epsilon_M \gam_1\gam_2$,
where $\epsilon_M$ is the coupling energy of the MZMs $\gamma_1$ and $\gamma_2$.
The Majorana operators are related to the regular complex fermion
through the transformation of $\gam_1=f+\fdg$ and $\gam_2=-i(f-\fdg)$.
Using the complex fermion representation,
the tunnel-coupling of the Majorana quantum wire to the two normal leads
in the three-terminal device is described as \cite{Li12}
\bea\label{Ham2}
H' = \sum_{\alpha=1,2}\sum_k t_{\alpha k}
\left[ (b^{\dagger}_{\alpha k}f
+(-1)^{\alpha+1} b^{\dagger}_{\alpha k}f^{\dagger})+{\rm H.c.}\right] .
\eea
$b^{\dagger}_{\alpha k}$ ($b_{\alpha k}$) are the creation (annihilation) operators
of electrons in the leads, while the leads are described by
$H_{\rm leads}= \sum_{\alpha=1,2}\sum_k \epsilon_{\alpha k} b^{\dagger}_{\alpha k} b_{\alpha k}$.
It should be noted that in $H'$ the tunneling terms only conserve charges modulo $2e$,
which actually correspond to the well known AR process.

Following Refs.\ \cite{Li05,Li14}, the tunnel-coupling Hamiltonian of \Eq{Ham2}
and the associated AR physics allow us to construct the MME as
\bea\label{ME-1}
\dot{\rho} &=& -\frac{i}{\hbar}[H_M,\rho]
+ \sum_{\alpha=1,2} \left(\Gamma_{\alpha}^{(+)} {\cal D}[f^{\dagger}]\rho
+ \Gamma_{\alpha}^{(-)} {\cal D}[f]\rho \right)   \nl
&& + \sum_{\alpha=1,2} \left(\widetilde{\Gamma}_{\alpha}^{(+)} {\cal D}[f]\rho
+ \widetilde{\Gamma}_{\alpha}^{(-)} {\cal D}[f^{\dagger}]\rho \right)  \,.
\eea
The Lindblad superoperator is defined by
${\cal D}[A]\rho= A\rho A^{\dagger}-\frac{1}{2}\{A^{\dagger}A, \rho\}$
and the rates in this {\it generalized} master equation are introduced as
\begin{subequations}
\bea\label{rate}
&& \Gamma_{\alpha}^{({\pm})}=\Gamma^e_{\alpha} N_{\alpha}^{(\pm)}\,, ~~
N_{\alpha}^{(\pm)}= \int d\omega n_{\alpha}^{(\pm)}(\omega)
\widetilde{\delta}(\omega-\epsilon_M) \,,  \nl
\\
&& \widetilde{\Gamma}_{\alpha}^{({\pm})}=\Gamma^h_{\alpha}
\widetilde{N}_{\alpha}^{(\pm)}\,, ~~
\widetilde{N}_{\alpha}^{(\pm)}= \int d\omega n_{\alpha}^{(\pm)}(\omega)
\widetilde{\delta}(\omega+\epsilon_M) \,. \nl
\eea
\end{subequations}
The superscripts ``$e$" and ``$h$" of the rates denote coupling of the quasiparticle
to the leads via ``electron" and ``hole" components, respectively.
We have also denoted the Fermi occupied function by $n^{(+)}_{\alpha}$
and the unoccupied function by $n^{(-)}_{\alpha}=1-n^{(+)}_{\alpha}$.
The spectral function $\widetilde{\delta}(\omega\mp\epsilon_M)$ is a generalization
from the Dirac $\delta$-function to a Lorentzian, which reads as
\bea\label{t-delt}
\widetilde{\delta}(\omega\mp\epsilon_M)
= \frac{1}{\pi}\frac{\Gamma}{(\omega\mp\epsilon_M)^2+\Gamma^2} \,,
\eea
where the broadening width is given by
$\Gamma=\sum_{\alpha} (\Gamma^e_{\alpha}+\Gamma^h_{\alpha})/2$.

We may have two remarks on the above MME.
{\it (i)}
The Lorentzian spectral function (instead of the Dirac-$\delta$ function)
properly accounts for the level broadening effect.
This generalization makes the MME applicable for transport under small bias voltage,
while it is well known that the usual Born-Markov-Lindblad master equation
is applicable only under large bias limit.
{\it (ii)}
The two Lindblad terms in the first round brackets in \Eq{ME-1}
are from the normal tunneling process,
while the two terms in the second round brackets from the Andreev process.
Accordingly, the conservation of energy is reflected differently
in the rate expressions, i.e., by the different centers
of the spectral functions $\widetilde{\delta}(\omega\mp \epsilon_M)$.

The MME can be straightforwardly solved using the basis of number-states
$\{|0\ra, |1\ra \}$ of the complex fermion $f$ (i.e., $n_f=0,1$).
In particular, for steady state, let us denote the density matrix as
$\bar{\rho}=p_0|0\ra\la 0|+p_1|1\ra\la 1|$.
The steady-state currents, e.g., the left-lead current,
which contains two components, $I_L=I^{(1)}_L+I^{(2)}_L$, can be calculated as
\bea\label{current-3}
I^{(1)}_L = \frac{e}{\hbar} [\Gamma^{(+)}_1 p_0 - \Gamma^{(-)}_1 p_1], ~
I^{(2)}_L = \frac{e}{\hbar} [\widetilde{\Gamma}^{(+)}_1 p_1
- \widetilde{\Gamma}^{(-)}_1 p_0].
\eea
Physically, $I^{(1)}_L$ is contributed by the conventional tunneling process
and $I^{(2)}_L$ is from the Andreev process (including also the CAR process).
More specifically, let us apply the above formal result
to the setup considered in Ref.\ \cite{Bee08},
where the two normal leads are equally biased with respect to the Fermi level
of the grounded superconductor, i.e., $\mu_L=\mu_R=eV$ and $\varepsilon_F=0$.
At zero temperature, we obtain
\bea\label{current-1}
I_L= \frac{e}{\hbar} \frac{\Gamma_1}{\pi}
\left[ \arctan(\frac{eV-\epsilon_M}{\Gamma})
+ \arctan(\frac{eV+\epsilon_M}{\Gamma} )   \right]  .
\eea
Here we have assumed $\Gamma^e_1=\Gamma^h_1\equiv \Gamma_1$.
In the following,
e.g., after \Eq{IL-Sm-1} and in the section ``{\it Local conductance}",
we will take this result --derived from the BdG-free master equation approach--
as a reference for comparisons between the two BdG treatments,
by employing the specific setup analyzed in Ref.\ \cite{Bee08}.
Before doing that, we first present
a modified BdG treatment within the scattering matrix formalism. \\
\\
{\it Modified BdG Treatment}.---
Following Refs.\ \cite{Bee08,Law09,Vu18,BCS-1,ABG02},
the scattering $S$ matrix has been formulated as
\bea\label{S-matrix}
S(\omega)= 1 - 2\pi i W^{\dagger} (\omega-H_M + i\pi W W^{\dagger})^{-1} W \,.
\eea
For transport through the MZMs, within the BdG formalism,
one can use either the Majorana modes $\{|\Phi_1\ra, |\Phi_2\ra\}$
or the eigenstates $\{|\Psi_+\ra, |\Psi_{-}\ra \}$,
to be coupled to the electron and hole components of the leads,
$\{|e_L\ra, |e_R\ra, |h_L\ra, |h_R\ra \}$.
As a result, the coupling operator $W$ is a $2\times 4$ matrix.
However, viewing that the negative-energy state $|\Psi_{-}\ra $
is nothing but the equivalent after removing an existing quasiparticle,
as a modified BdG treatment, we propose to keep only $|\Psi_+\ra$ to couple to
the electron and hole states of the leads,     
with thus a $1\times 4$ coupling matrix given by
\bea\label{W-matrix}
W = (t_l u_l,\, t_r u_r,\, -t_l v_l^*,\, - t_r v_r^*) \,,
\eea
where $u_{l(r)}$ and $v_{l(r)}$ are, respectively, the electron and hole amplitudes
of $|\Psi_+\ra$ at the left (right) end of the wire.

We emphasize that only this modified treatment
(keeping only the positive-energy state $|\Psi_{+}\ra $)
can give consistent result with the MME based on the number-state description.
In other words, one should not treat the negative energy state $|\Psi_{-}\ra $
as real quasiparticle excitation to participate in the charge transport dynamics.
Actually, its superposition with the positive eigenstate $|\Psi_{+}\ra $
is the reason that results in the vanishing transmission/teleportation and crossed AR
when $\epsilon_M\to 0$ \cite{Bee08,Zoch13}, as analyzed in detail
based on the simple ``dot-wire-dot" model system \cite{Li20}.     

Inserting \Eq{W-matrix} into (\ref{S-matrix}), we obtain
\bea\label{modified-S}
&& S=1-2\pi i  \nu z^{-1} \nl
&&
\times \left(
\begin{array}{cccc}
|t_l|^2|u_l|^2 &  t_l^* t_r u_l^*u_r & -|t_l|^2u_l^*v_l^* & -t_l^* t_r u_l^*v_r^* \\
t_l t_r^* u_l u_r^* &  |t_r|^2|u_r|^2 &-t_l t_r^* v_l^*u_r^* & -|t_r|^2u_r^*v_r^*  \\
-|t_l|^2 u_l^* v_l &  -t_l^* t_r v_l^*u_r &|t_l|^2|v_l|^2 & t_l^* t_r v_l v_r^* \\
-t_l t_r^* u_lv_r  & -|t_r|^2u_rv_r & t_l t_r^* v_l^* v_r & |t_r|^2|v_r|^2
\end{array}
\right)  \nl
\eea
We have introduced $\nu$ for the density-of-states of the leads,
and $z=(\omega-\epsilon_M)+i\Gamma$.
The total coupling rate $\Gamma$ is the same as defined in the MME,
while more explicitly we have
$\Gamma^e_{\alpha}=2\pi\nu |t_{\alpha}|^2 |u_{\alpha}|^2$
and $\Gamma^h_{\alpha}=2\pi\nu |t_{\alpha}|^2 |v_{\alpha}|^2$.
Here the index ``$\alpha$" in $\{t_{\alpha}, u_{\alpha}, v_{\alpha}\}$
also corresponds to the left (``$l$") and right  (``$r$") sides
(for $\alpha=1$ and $2$, respectively).
In the ideal case $\epsilon_M=0$, we have $|u_{\alpha}|^2 = |v_{\alpha}|^2$,
thus $\Gamma^e_{\alpha}=\Gamma^h_{\alpha}\equiv \Gamma_{\alpha}$.
Based on the result of the $S$ matrix, one can obtain the various
transport coefficients, such as
${\cal T}^{eh}_{11A}=|s_{13}|^2 = \Gamma^e_1\Gamma^h_1/|z|^2$ for the local AR,
${\cal T}^{eh}_{12A}=|s_{14}|^2 = \Gamma^e_1\Gamma^h_2/|z|^2$ for the crossed AR,
and ${\cal T}^{ee}_{12}=|s_{12}|^2 = \Gamma^e_1\Gamma^e_2/|z|^2$
for the electron transmission/teleportation.
Further, the various currents, e.g., the left-lead current, can be obtained as
\bea\label{IL-Sm-1}
I_L = \frac{2e}{h} \int^{eV}_{-eV} d\omega \,
[{\cal T}^{eh}_{11A}(\omega)+ {\cal T}^{eh}_{12A}(\omega)] \,.
\eea
One can easily check that this gives precisely the same result of \Eq{current-1}.

\begin{figure}[h]
\includegraphics[scale=0.6]{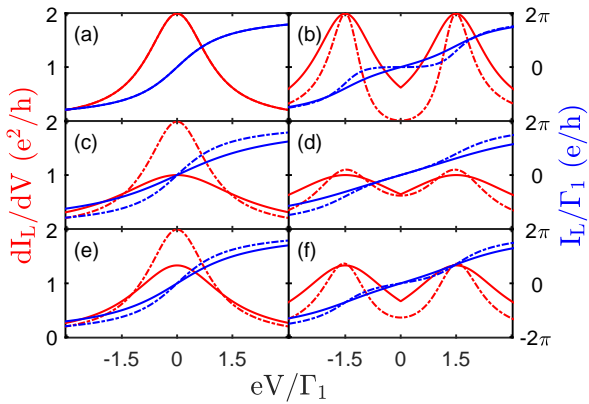}
\caption{
Conductances (red curves) and currents (blue curves)
mediated by the MZMs in a two-lead setup
with coupling rates $\Gamma_1$ and $\Gamma_2$.
The coupling asymmetry is characterized by $\eta=\Gamma_2/\Gamma_1$,
while the results for $\eta=0$, 1, and 0.5
are shown in (a) and (b), (c) and (d), and (e) and (f), respectively.
In the left panels (a), (c) and (e), we consider $\epsilon_M=0$;
while in the right panels (b), (d) and (f), we assume $\epsilon_M=1.5\Gamma_1$.
In the whole plots, the solid curves display results from the modified BdG treatment,
and the dashed curves from the standard BdG method \cite{Bee08,BCS-1}. }
\end{figure}

Importantly, in the result of \Eq{IL-Sm-1}, the CAR contribution ${\cal T}^{eh}_{12A}$
is nonzero even at the limit $\epsilon_M\to 0$.
Also, since ${\cal T}^{eh}_{12A}={\cal T}^{ee}_{12}$
($\Gamma^e_{\alpha}=\Gamma^h_{\alpha}$ under $\epsilon_M\to 0$),
the teleportation channel is not closed
even if the two MZMs ($\gamma_1$ and $\gamma_2$) have no coupling.      \\
\\
{\it Local Conductance}.---
Let us consider first the simplest {\it single-lead device}
where the probe lead is tunnel-coupled to
the grounded Majorana wire from one side (e.g., the left side).
We can use either the current formula based on the $S$ matrix
or the formula based on the master equation,
both giving the same results provided the modified BdG treatment is applied.
Using \Eq{current-3},
we may split the total current of \Eq{current-1}  into two parts
\bea\label{current-5}
I^{(1,2)}_{L}= \frac{e}{\hbar}\int^{\mu_1}_{\mu_2} d\omega
 \left(\frac{\Gamma^e_1\Gamma^h_1}{\Gamma^e_1+\Gamma^h_1}\right)
 \widetilde{\delta}(\omega\mp\epsilon_M) \,.
\eea
Here we introduced $\mu_1=-\mu_2=eV$.
The spectral function $\widetilde{\delta}(\omega\mp\epsilon_M)$
takes the same form of \Eq{t-delt}, while for single-lead device
the Lorentzian width is reduced to $\Gamma=(\Gamma^e_1+\Gamma^h_1)/2$.
Then, the differential conductance can be computed through
\bea\label{G-1}
G=e\left( \frac{\partial I^{(1)}_L}{\partial \mu_1}
+ \frac{\partial I^{(2)}_L}{\partial \mu_2} \right)|_{\mu_1=eV,\, \mu_2=-eV} \,.
\eea
From \Eqs{current-5} and (\ref{G-1}), we obtain the well-known Majorana conductance
$G=\frac{e^2}{h}(1+1)=\frac{2e^2}{h}$,
i.e., the quantized ZBP at $eV=\epsilon_M\to 0$
\cite{Law09,BNK11,Flen10,Flen16,Kou12,Kou19},
which holds also for the Majorana-induced resonant AR conductance
at $eV=\epsilon_M\neq 0$ \cite{Law09}.
We notice that, for the single-lead setup, both BdG treatments predict the same height
of Majorana conductance peak, $2e^2/h$, as shown in Fig.\ 1(a) and (b),
despite that the conventional BdG treatment
predicts a narrower width of the conductance peak when $\epsilon_M\neq 0$.

Next, let us consider the {\it two-lead device} setup,
which will reveal more remarkable differences between the two BdG treatments.   
Formally, the current formula is the same as \Eq{current-5}, needing only
by adding the CAR contribution such as
$\Gamma^e_1\Gamma^h_2/(\Gamma^e_1+\Gamma^h_2)$.
Moreover, the Lorentzian width in $\widetilde{\delta}(\omega\mp \epsilon_M)$
is now given by $\Gamma=(\Gamma^e_1+\Gamma^h_1+\Gamma^e_2+\Gamma^h_2)/2$.
In addition to increasing the resonance width,
this more coupling to the right lead would decrease the heights
of the various transmission coefficients, such as the LAR and CAR coefficients
(${\cal T}^{eh}_{11A}$ and ${\cal T}^{eh}_{12A}$) as $\omega\to\epsilon_M$.
We may term this type of consequences as a
Majorana nonlocal-coupling-effect on the self energy.  
In particular, for symmetric coupling,
the four coupling rates can be considered identical,
and the heights of the the LAR and CAR peaks are reduced to 1/4,
for either $\epsilon_M=0$ or not.
At the limit $\epsilon_M\to 0$,
owing to the complete ``disconnection" between $\gamma_1$ and $\gamma_2$,
the conventional BdG treatment predicts that
${\cal T}^{eh}_{12A}=0$ and ${\cal T}^{eh}_{11A}|_{\omega\to\epsilon_M}=1$, respectively.
Then, the ZBP of the LAR conductance (in the left lead)
would remain the same height as $2e^2/h$,
being unaffected by the Majorana coupling to the opposite (right) lead.
In contrast, based on either the modified BdG treatment or the MME,
we predict the ZBP height as
$G=\frac{2e^2}{h}\times \frac{1}{4}+ \frac{e^2}{h}(\frac{1}{4}+\frac{1}{4})=e^2/h$.
Here, the first part is from the LAR contribution,
while the second part from the CAR process.
In Fig.\ 1(c) and (d), we show the full results of this symmetric two-lead device,
for both $\epsilon_M=0$ and $\epsilon_M\neq 0$.
We find that the former case reveals greater difference between the two treatments.
In Fig.\ 1(e) and (f), we also show the results for asymmetric coupling.
Big difference exists as well in this case, particularly for $\epsilon_M\to 0$:
the conventional BdG treatment predicts a constant ZBP of $2e^2/h$;
while the modified BdG treatment predicts that
the other side coupling will affect the height of the ZBP,
e.g., for $\Gamma_2=\Gamma_1/2$, which is $1.5 e^2/h$.    \\    
\\
{\it Teleportation Conductance}.---
Now we turn to the unequally biased two-lead device.
For this setup ($\mu_L\neq\mu_R$), in addition to the AR process, the tunneling of electron
between the two leads has contribution to the current.
Again, applying \Eq{current-3}, we obtain (at zero temperature)
\bea\label{current-4a}
I_{\alpha} = \frac{2e}{h}  \Gamma_{\alpha}
\left[ \arctan(\frac{\mu_{\alpha}-\epsilon_M}{\Gamma})
+\arctan(\frac{\mu_{\alpha}+\epsilon_M}{\Gamma}) \right] .
\eea
Here we have assumed the convention that
$I_{1,2} = I_{L,R}$ and $\mu_{1,2}= \mu_{L,R}$.
Rather than the total current, based on \Eq{current-3},
simple analysis also allows us to know
the individual components in $I^{(1)}_L$ and $I^{(2)}_L$, which result in
\bea\label{current-4}
&&I_L = \frac{2e}{h}\int^{\mu_L}_{-\mu_L} d\omega\, {\cal T}^{eh}_{11A}(\omega)  \nl
&&~ + \frac{e}{h}\left( \int^{\mu_L}_{-\mu_R} d\omega\, {\cal T}^{eh}_{12A}(\omega)
+\int^{\mu_R}_{-\mu_L} d\omega\, {\cal T}^{eh}_{21A}(\omega)  \right)  \nl
&&~
+ \frac{e}{h}\left( \int^{\mu_L}_{\mu_R} d\omega\, {\cal T}^{ee}_{12}(\omega)
+\int^{-\mu_R}_{-\mu_L} d\omega\, {\cal T}^{hh}_{21}(\omega)  \right)  .
\eea
The right-lead current $I_R$ can be similarly decomposed. Notice that, here,
besides the contribution from the LAR (result of the first line)
and CAR (result of the second line),
the third line is the current from the left to the right lead
through the teleportation channel.
That is, the first term of the third line is from the
electron-to-electron tunneling (from the left to the right lead),
while the second term corresponds to an equivalent
hole-to-hole tunneling (from the right to the left lead).
In more detail, as an example, the second term of the third line
was derived from $I^{(2)}_L (1h\leftarrow 2h) =\frac{e}{\hbar}
(\widetilde{\Gamma}^{(+)}_1 \widetilde{\Gamma}^{(-)}_2
- \widetilde{\Gamma}^{(-)}_1 \widetilde{\Gamma}^{(+)}_2)/2\Gamma$,
where ``$2h$" stands for the hole in the right lead,
and ``$1h$" the hole in the left lead.

One can check that, based on the modified BdG $S$-matrix solution \Eq{modified-S},
the sum of all terms in \Eq{current-4} precisely recovers the result of \Eq{current-4a}.
However, rather than the total current, below we are more interested
in the current component through the teleportation channel,
i.e., the third line of \Eq{current-4}.
For this purpose, we propose to extract this part of current via
the consideration $\Delta I_L=I_L-\widetilde{I}_L$,
where $\widetilde{I}_L$ denotes the sum of the AR currents
(both LAR and CAR -- the first and second lines of \Eq{current-4}),
which flows back from the grounded superconductor to the left lead
and can be measured as a branch circuit current.
Then, from the third line of \Eq{current-4},
we further obtain the differential conductance
(termed as {\it teleportation conductance} in this work)
\bea\label{dGLL}
\Delta G_{LL}=\frac{d(\Delta I_L)}{d V_L}
= \frac{e^2}{h} \left[ {\cal T}^{ee}_{12}(\mu_L) + {\cal T}^{hh}_{21}(-\mu_L)  \right] \,.
\eea
Based on the $S$-matrix solution of the modified BdG treatment,
\Eq{modified-S}, we have
${\cal T}^{ee}_{12}(\omega)=\Gamma^e_1 \Gamma^e_2 / |z|^2$
and ${\cal T}^{hh}_{21}(\omega)=\Gamma^h_2 \Gamma^h_1 / |z|^2$,
where $|z|^2=(\omega-\epsilon_M)^2+\Gamma^2$.
To be more specific, we may assume the bias voltage as $\mu_R=0$ and $\mu_L=eV_L>0$.
From the above result, it becomes clear that as $\epsilon_M\to 0$
the teleportation current $\Delta I_L$
and the differential conductance $\Delta G_{LL}$ are {\it nonzero}.
This is a very important result, which indicates that,
even at the limit $\epsilon_M\to 0$, the teleportation channel is still open.   

Following Ref.\ \cite{BCS-1} as an example, which generalizes Ref.\ \cite{Bee08}
by considering $\Gamma^e_{\alpha}\neq \Gamma^h_{\alpha}$ when $\epsilon_M\neq 0$,
the conventional BdG treatment yields the solution of $S$ matrix which gives
\bea\label{T12ee-mBdG}
{\cal T}^{ee(hh)}_{12(21)} = |(\omega+i \Gamma)\xi \pm \epsilon_M\Gamma|^2/|z|^2 \,,
\eea
where $|z|^2=(\omega^2-\epsilon^2_M-\Gamma^2)^2 + 4\omega^2\Gamma^2$.
Here we introduced $\xi=\Gamma^e-\Gamma^h$
and, for the sake of simplicity,
assumed that $\Gamma^e_1=\Gamma^e_2\equiv\Gamma^e$,
$\Gamma^h_1=\Gamma^h_2\equiv\Gamma^h$, and $\Gamma=\Gamma^e + \Gamma^h$.
Substituting \Eq{T12ee-mBdG} into \Eq{dGLL}, we obtain
\bea\label{dGLL-2}
\Delta G_{LL}= (\frac{2e^2}{h}) \,
\frac{(\xi\mu_L + \epsilon_M\Gamma)^2+\xi^2\Gamma^2}{|z|^2}  \,.
\eea
Here, in $|z|^2$, one should take $\omega=\mu_L$.
We may emphasize that this result predicts
that the teleportation channel vanishes when $\epsilon_M\to 0$.   
In this context,
one may notice that $\xi_{\alpha}\equiv \Gamma^e_{\alpha}-\Gamma^h_{\alpha}
=2\pi\nu |t_{\alpha}|^2(|u_{\alpha}|^2-|v_{\alpha}|^2)$,
which is closely related to the so-called local BCS charges \cite{BCS-1a,BCS-1,BCS-2},
$q_{\alpha}=|u_{\alpha}|^2-|v_{\alpha}|^2$.
Moreover, our numerical simulation based on the Kitaev lattice model \cite{Kita01}
reveals that $q_{\alpha}/|u_{\alpha}|^2 \propto \epsilon_M$,
in the regime of relatively small Majorana coupling energies.
Therefore, for the symmetric case $\xi_1=\xi_2=\xi$,
we may denote $\xi/\Gamma = K\epsilon_M$ and reexpress the conductance
under the conditions $V_L\to 0$ and $\epsilon_M \ll \Gamma$, as
\bea\label{dGLL-3}
\Delta G_{LL}= (\frac{2e^2}{h}) \,
(K^2 + 1/\Gamma^2 )\, \epsilon^2_M   \,.
\eea
In Fig.\ 2, based on simulation of the Kitaev model
for a spinless $p$-wave superconductor \cite{Kita01},
we display a logarithmic plot for this conductance as a function of $\epsilon_M$,
to demonstrate the qualitative scaling behavior of $\Delta G_{LL}\propto \epsilon^2_M$,
by noting that the prefactor
$\widetilde{K}^2=K^2+1/\Gamma^2$ only depends on $\epsilon_M$ weakly.
The weak dependence is originated from the coupling rate $\Gamma$
which decreases with increasing $\epsilon_M$,
owing to the wavefunction extension of the Majorana bound states
(towards the inner part of the quantum wire),
while $K$ keeps almost a constant.
Therefore, as a consequence of the approximate $\epsilon^2_M$-scaling behavior,
the conventional BdG treatment predicts that
the teleportation channel will be closed, when $\epsilon_M\to 0$.
However, as shown by the blue-solid-line in Fig.\ 2,
the modified BdG treatment predicts that
the teleportation channel remains open even at the limit $\epsilon_M\to 0$,
and the teleportation conductance $\Delta G_{LL}$ is almost independent of $\epsilon_M$.  \\  
\\
{\it Summary}.---
We have constructed a Majorana master equation
which only associates with the BdG free occupation-number-states of the regular fermion.
The results from the master equation approach forced us to modify the BdG treatment,
in order to achieve consistent results.
For the two-lead (three-terminal) transport setup,
we revealed the existence of nonvanishing channels
of teleportation and crossed Andreev reflections even if $\epsilon_M\to 0$.
We also predicted different heights of the zero-bias-peak of the local conductance
and different $\epsilon_M$-scaling behaviors of the teleportation conductance.
Verification of the predictions by experiments will be of great interest.

\begin{figure}[h]
\includegraphics[scale=0.55]{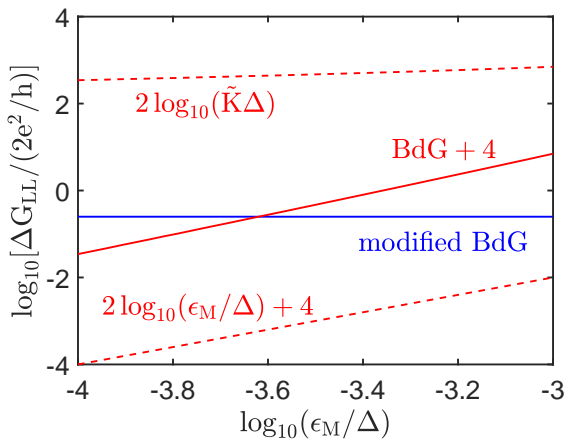}
\caption{
Scaling behavior of the teleportation conductance $\Delta G_{LL}$
with the Majorana coupling energy $\epsilon_M$,
from simulation based on the Kitaev model \cite{Kita01},
$ H_M = \sum^{N}_{j=1} [-\mu c^{\dg}_{j}c_j - t  (c^{\dg}_{j}c_{j+1}+{\rm h.c.})
 + \Delta (c_{j}c_{j+1}+{\rm h.c.})]$,
where $\mu$ is the chemical potential, $t$ is the hopping energy,
and $\Delta$ is the superconducting order parameter.
In numerical simulations, we set $t=\Delta=1.0$ and vary $\mu$
to realize the change of $\epsilon_M$, under the condition
$\epsilon_M \ll \Gamma$ ($\Gamma$ is the coupling rate to the leads
in the symmetric case as we assumed).
We also consider the zero-bias limit $\mu_L\to 0$
while always setting $\mu_R=\epsilon_F=0$
($\epsilon_F$ is the chemical potential of the superconductor).
{\it (i)}
The result shown by the blue-solid-line is from the modified BdG treatment,
which reads as $\Delta G_{LL}=e^2/2h$
(in the symmetric coupling case).
{\it (ii)}
The result depicted by the red-solid line is from the conventional BdG treatment,
which indicates an approximate scaling behavior of $\Delta G_{LL}\sim \epsilon^2_M$.
Based on $\Delta G_{LL} = (\frac{2e^2}{h})\widetilde{K}^2\epsilon^2_M $,
we also plot the two multiplying factors separately, by the red-dashed lines.   }
\end{figure}

\vspace{0.5cm}
{\flushleft\it Acknowledgements.}---
This work was supported by the
National Key Research and Development Program of China
(No.\ 2017YFA0303304) and the NNSF of China (Nos.\ 11675016, 11974011 \& 61905174).

\end{document}